\documentclass{ws-p8-50x6-00}
\def\bea{\begin{eqnarray}}
\def\eea{\end{eqnarray}}

\begin{document}
\vsize17.0cm
\newcommand{\insertplotg}[1]{
\begin{center}\leavevmode\epsfysize=2.5in \epsfbox{#1}\end{center}}

\title{Hyperon ratios at RHIC and the coalescence predictions 
	at mid-rapidity }
\author{J. Zim\'anyi, P. L\'evai, T. Cs\"org\H o, T.S. Bir\'o
	}
\address{MTA KFKI RMKI, H-1525 Budapest 114, POB 49, Hungary \\
	e-mail: jzimanyi@sunserv.kfki.hu
	}

\maketitle

\abstracts{
Quark coalescence predictions for various multi-strange
baryon to anti-baryon ratios for central Au+Au collisions at RHIC
energies are compared to preliminary data of the STAR collaboration.
The formation of Quark Matter and the sudden 
recombination of its constituent quarks 
into hadrons is found to be in agreement with these preliminary data.
It seems that strange hadron and antihadron production in Au+Au collisions
 at RHIC is similar to that in Pb+Pb collisions at CERN SPS.
}

In the  Quark Matter'99 conference the possibility of an
interesting scenario for the RHIC reactions was
pointed out.\cite{QM99} Namely we may meet
 the situation that in the first stage of the collision a
 sort of quark gluon plasma fireball is formed. This fireball expands
and cools, and just before the hadronization this matter may
change into a Quark Matter similar to that
  state which was present at the SPS Pb+Pb
collisions at the beginning of hadronization. The basic degrees of 
freedom of this state are the dressed valence quarks, the gluonic
degrees of freedom are suppressed e.g. due to their large effective 
mass.\cite{LevHein} The earlier fragmentation of
these gluons resulted in new valence quarks, that are the constituents 
of the final state hadrons. Keeping in mind
this possibility, theoretical predictions were made
for the RHIC Au+Au collisions based on a coalescence
 model, which succesfully described - among others - the antihyperon
to hyperon ratios at SPS.

Recently preliminary experimental data were shown for these
ratios by the STAR collaboration.\cite{STAR,STAR2} Thus it is timely
to compare these new experimental results with the
 theoretical predictions.

\section{The Linear Coalescence Model}

In the following we shall demonstrate the deficiency of
the {\bf linear} coalescence model in a very simple case.
The idea of coalescence was created many years ago for the understanding 
of some low energy nuclear physics problems. 
Among others the production of deuterons 
in low energy heavy ion reaction was explained by the assumption 
that a neutron and a proton will coalesce if they are near to each other 
in space and momentum space.
 We shall denote by $k$ the {\it number of particles of type $k$ within 
the rapidity interval, in which the particle ratios are 
constant}.  (In low energy reactions these numbers are
 the total numbers of particles within the fireball.)
 The number of produced deuterons, $d$,
is assumed to be proportional to the number of protons, $p$, and 
the number of neutrons, $n$. The factor of proportionality, $c_{\rm d}$,
can be calculated from the elementary cross sections and
the phase-space distribution of protons and neutrons.
This leads to the linear coalescence model prediction
\be
        d = c_{\rm d} \, p \, n.   \label{e:lin_co}
\ee
This kind of relationship expresses 
 the essence of the linear coalescence model. It  is
 an acceptable approximation as long as a small fraction of protons 
and neutrons will coalesce into deuterons, $c_{\rm d} << 1$. 
The linear coalescence model, however, has a serious deficiency 
if a substantial part of the nucleons are forced into deuterons. 
To demonstrate the problem, 
let us assume, that $p=n=A$, and 
all the nucleons have  to be confined into deuterons at the end
of the fireball lifetime. This way we
model a situation similar to
the hadronization of the constituent quark matter. 
The conservation of baryon number demands that the 
baryon charge should be preserved during the coalescence process:
\be
p + n =  2 \, d \label{e:constrain} 
\ee
which leads to the requirement
\be 
p + n = 2 \, c_{\rm d} \, p \, n,
\label{nco}
\ee 
that implies 
\be 
	2 A  =   2 c_{\rm d} A^2 .  \label{e:lin}
\ee

Baryon number conservation and linear coalescence thus requires 
\be
    c_{\rm d}  = 1 /A . \label{e:wrong}
\ee
However, the coalescence coefficient $c_{\rm d}$ can be calculated 
from the cross section of the $p+n\rightarrow d$ reaction and the 
phase-space distribution of protons and neutrons, independently
of the total number of these particles, $2A$.  
Hence eq.~(\ref{e:wrong}) cannot be fullfilled in the general case
and the assumption of linear coalescence leads to contradiction.

What is the logical error in the linear coalescence model
which leads to the above, obviously unacceptable result? 
Essentially this is the lack of a feed-back:  those
nucleons, who are already bound in a deuteron, cannot be
used any further for the formation of new deuterons.
Out of $A$ protons and $A$ neutrons, exactly $A$ deuterons can
be formed,
if a bound state formation happens with unit probability.  
However, linear coalescence calculation would predict 
$\propto A^2$ deuterons,
which is way too large, if $A>>1$.

This sort of error in linear coalescence can be 
corrected by introducing a suitable
renormalization factor, $b$, which takes care of the
additional reduction of produced particles due to the lack
in the reactivity of the already bound constituents.
This leads to reconciliation of the constraints from conservation laws
and the proportionality of produced particles with the available number
of their constituents.

With this re-normalization factor, the coalescence equation  
eq.~(\ref{e:lin_co})
is modified, 
\be
        d  =  c_{\rm d}\ b_{\rm p} p \  b_{\rm n} n \ , \label{e:nco}
\ee
and eq.(\ref{nco}) still ensures the conservation of the
baryon charge.  In the simple case
of $p=n=A$, this determines the normalization factor, 
$b_{\rm p} = b_{\rm n} = b$
and predicts the of deuterons, $d$, as follows:
\bea
    b & = & \frac{1}{ (c_{\rm d} A)^{1/2} }, \\
    d & = & A \ne p \, n = A^2.
\eea
Note also that the total number of deuterons after the 
completion of the recombination becomes independent from the 
exact value of the coalescence coefficient $c_{\rm d}$,
because of  the constraint that all protons and neutrons have to
be converted to deuterons in this example. This has to be contrasted
to the result of linear coalescence calculations, where the
formation of a small number of deuterons from a gas of protons
and neutrons depends essentially on the coefficient of coalescence,
$c_{\rm d}$.

The method of introduction of normalization coefficients
 can be generalized for more complex cases that
involve the full recombination of different type of
 constituents into  various kind of bound states
subjected to conservation laws. In case of hadronization
(recombination
of valence quarks into hadrons), the resulting 
non-linear quark coalescence model is the  ALCOR 
model,\cite{ALCOR95,SQM97,ALCOR99} 
the name  of which
 stands for ALgebraic COalescence in Rehadronization.

\section{ALCOR:  Quark Combinatorics in Rehadronization}

The nonlinear ALCOR coalescence model was created for situations, 
where the subprocesses are not independent, they compete with
each other. In this model the coalescence equations relating the number 
of a given type of hadron to the product of the numbers of 
different quarks from which the hadron consists reads as: 

\begin{eqnarray}
    {p}  = C_{\rm p} \,  b_{\rm q} \,  b_{\rm q} 
    \,  b_{\rm q} \,  q \,  q \,  q
&\ \ & {\overline p}  = C_{\overline p} \,  
b_{\overline {\rm q}} \,  b_{\overline {\rm q}} \,  
b_{\overline {\rm q}} \, 
{\overline q} \,  {\overline q} \,  {\overline q} 
\nonumber \\
    {\Lambda} = C_{\Lambda} \,  
    b_{\rm q} \,  b_{\rm q} \,  b_{\rm s} \,  q \,  q \,  s
&\ \ & {\overline \Lambda}  = C_{\overline \Lambda} \,  
b_{\overline {\rm q}} \,  b_{\overline {\rm q}} \,  
b_{\overline {\rm s}} \, 
{\overline q} \,  {\overline q} \,  {\overline s} 
\nonumber \\
    {\Xi}  = C_{\Xi} \,  
    b_{\rm q} \,  b_{\rm s} \,  b_{\rm s} \,  q \,  s \,  s
&\ \ & {\overline \Xi}  = C_{\overline \Xi} \,  
b_{\overline {\rm q}} \,  b_{\overline {\rm s}} \,  
b_{\overline {\rm s}} \, 
{\overline q} \,  {\overline s} \,  {\overline s} 
\nonumber \\
    {\Omega}  = C_{\Omega} \,  
    b_{\rm s} \,  b_{\rm s} \,  b_{\rm s} \,  s \,  s \,  s
&\ \ & {\overline \Omega}  = C_{\overline \Omega} \,  
b_{\overline {\rm s}} \,  b_{\overline {\rm s}} \,  
b_{\overline {\rm s}} \, 
{\overline s} \,  {\overline s} \,  {\overline s} 
\label{coal}
\end{eqnarray}

\begin{eqnarray}
    \pi  &=& C_\pi \,  
b_{\rm q} \,  b_{\overline {\rm q}} \,  q \,  {\overline q}
\nonumber \\
    K  &=& C_K \,  
b_{\rm q} \,  b_{\overline {\rm s}} \,  q \,  {\overline s}
\nonumber \\
    {\overline K}  &=& C_{\overline K} \,  
b_{\overline {\rm q}} \,  b_{\rm s} \,  {\overline q} \,  s
\nonumber \\
    {\eta} &=& C_{\eta} \,  
b_{\overline {\rm s}} \,  b_{\rm s} \,  {\overline s} \,  s
\label{coalm}
\end{eqnarray}

Here the normalization coefficients, $b_{\rm i}$,  are determined uniquely
by the requirement, that { the number of the constituent
quarks do not change during the hadronization ---
which is the basic assumption for all quark counting methods}:
\begin{eqnarray}
 s &=& 3 \,  \Omega + 2 \,  \Xi +  \Lambda +
       {\overline K} +  {\eta}  \nonumber \\
 {\overline s}  &=& 3 \,  {\overline \Omega} + 
2 \,  {\overline \Xi} +   {\overline \Lambda} +
        {K} +  {\eta} \nonumber \\
 q &=& 3 \,  p +  \Xi + 2 \,  \Lambda +  {K}+
       \pi  \nonumber \\
 {\overline q}  &=& 3 \,  {\overline p} + 
  {\overline \Xi} + 2 \,  {\overline \Lambda} +
        {\overline K}
    +  \pi  \ . \label{cons} 
\end{eqnarray}

In eq.~(\ref{cons}) $\pi $ is the number of directly produced pions.
(Most of the observed pions are created in the decay
of resonances.)
Substituting eqs.~(\ref{coal}-\ref{coalm}) into
 eq.~(\ref{cons}) one obtains  equations
for the normalization constants. These constants
are then given in terms of  quark numbers and  $ C_{\rm i} $ factors.

\section {ALCOR Predictions Compared to  Preliminary RHIC Data } 

\subsection{Parameter independent hyperon ratios for SPS and RHIC}
In order to arrive at
the detailed predictions of the ALCOR model, one has to
solve numerically the set of equations eqs.~(\ref{coal}-\ref{cons}).  
The ALCOR predictions for the particle multiplicities (especially 
for the antihyperon to hyperon ratios), were shown in 
Refs.{\cite{ALCOR95,SQM97}} for the SPS energy.

Later on an interesting 
observation was pointed out in Ref.\cite{Bial98} Namely
it was shown, that within the framework of the linear coalescence model 
in the antiparticle/particle ratios the uncertain 
$c_{\rm i}$ coalescence factors drop out, and thus one obtains direct 
relations between particle ratios. Unfortunately the linear 
coalescence model contradiscts to the different conservation laws.

In the later development, however, it was shown in Refs.~{\cite{QM99,Zim99}}
that even in the charge conserving normalized 
(non-linear) coalescence model one can obtain 
parameter independent ratios between particle 
multiplicities.  These relations have the form:
 
\begin{equation}
\frac{\overline \Lambda}{\Lambda} = D \,  \frac{\overline p}{p}\ \ ,
\ \ \ \ \
\frac{\overline \Xi}{\Xi} = D \,  \frac{\overline \Lambda}{\Lambda}\ \ ,
\ \ \ \ \
\frac{\overline \Omega}{\Omega} = D \,  \frac{\overline \Xi}{\Xi}\ \ ,
\label{ratios}
\end{equation}
where
\begin{equation}
 D=\frac{K^+}{K^-} \ \ .
\label{ratiosk}
\end{equation} 

Thus if the dynamics of hadronization is the coalescence process,
then these relations should hold among the measured quantities.
The preliminary results of the STAR experiment~\cite{STAR,STAR2}
 for Au+Au collision at $\sqrt{s}=130$ AGeV, 
\begin{eqnarray} 
 {\overline p} / {p}&=& 0.61 \pm 0.06  \nonumber \\[0.7ex]
{\overline \Lambda} / {\Lambda}&=& 0.73 \pm 0.03 \nonumber \\[0.7ex]
{\overline \Xi} / {\Xi}&=& 0.82 \pm 0.08  \nonumber \\[0.7ex]
{K^+} / {K^-}&=& 1.12 \pm 0.06
\end{eqnarray}
clearly satisfy the relations 
~eqs.(\ref{ratios}-\ref{ratiosk}), as required by quark combinatorics.

Similar results support the quark coalescence mechanism,
observed in Pb+Pb collision at $E_{\rm beam}=158$ AGeV energy
at CERN SPS:\cite{EXP1,EXP2}
\begin{eqnarray}
 {\overline p}/{p}&=& 0.07 \pm 0.01  \nonumber \\
{\overline \Lambda} / {\Lambda}&=& 0.133 \pm 0.007 \nonumber \\
{\overline \Xi} / {\Xi}&=& 0.249 \pm 0.019 \nonumber \\
{\overline \Omega} / {\Omega}&=& 0.383 \pm 0.081 
\label{SPSrat}
\end{eqnarray}
which also satisfy the relations among the 
ratios demanded by quark coalescence 
with
\begin{equation}
{K^+} / {K^-}= 1.8 \pm 0.2 \, .
\label{SPSratk}
\end{equation}

Hence the multi-strange baryon to anti-baryon ratios
in central Au+Au collisions at RHIC and in central
Pb+Pb collisions at CERN SPS are consistent with
the formation and the sudden rehadronization
of Quark Matter, consisting of valence
quarks and anti-quarks.

\subsection { Detailed  ALCOR  predictions for RHIC} 

The first ALCOR predictions for RHIC  were given
for Au+Au collision at $\sqrt{s}=200 \  GeV$ in Ref.\cite{QM99} 
The first data sets were taken at the bombarding energy
of $\sqrt{s}=130$ A GeV, and the preliminary values for
particle ratios were reported by the STAR Collaboration
in October 2000 at the XXXth International Symposium 
on Multiparticle Dynamics~\cite{STAR}
as well as at the Quark Matter in January 2001.\cite{STAR2}.
These data were similar to the ALCOR predictions for particle
ratios at $\sqrt{s}= 200$ AGeV.
Here we re-evaluate the ALCOR predictions for the lower than
expected values of $\sqrt{s}$.
We utilize the capability of the ALCOR model to work in a
reasonably small  rapidity bin ($\Delta y \approx 1$)
where all the available quarks and antiquarks are able to interact
to  form the measured mesons and baryons.
The calculation  of hadron multiplicities in the ALCOR 
model is based on the number of quarks and
antiquarks in the rapidity interval under study. 
These quark and antiquark numbers consist 
 of the $u{\overline u}$,
$d{\overline d}$, $s{\overline s}$ pairs produced  in the collision
and the $u$ and $d$ quarks of the colliding nuclei.
We utilize the measured (preliminary) value of the
central rapidity density of the negatively charged hadrons:\cite{STAR2}
\begin{equation}
{\rm d}N_{{\rm h}^-}/{\rm d}y = 264 \pm 18 \ \ .
\end{equation}
If we apply the parameter of the strangeness 
($f_{\rm s}=N_{{\rm s}{\overline {\rm s}}}/(N_{{\rm u}{\overline {\rm u}}}
+ N_{ {\rm d}{\overline {\rm d}}})=0.22$) and baryon
production ($\alpha_{\rm eff}=0.85$) determined at CERN SPS 
energy,\cite{ALCOR99} then we are very close to make a full
calculation of all hadronic
ratios. The only missing information is the number of stopped
nucleons in this rapidity region. (In previous calculations
at CERN SPS energy we have not used
 this information, because total hadron numbers were calculated
including all stopped nucleons.)
Choosing 197 newly produced $u{\overline u}$ and $d{\overline d}$
pairs (this means 86 $s{\overline s}$ pairs) and assuming  that 5 \% 
of the total nucleon numbers
stopped into the central unit of rapidity (which means 8 proton and
12 neutron), we can calculate the measured  rapidity densities and
hadronic ratios for the Au+Au collision at $\sqrt{s}=130$ AGeV
in the central rapidity as displayed in Table 1.

\begin{table}
\begin{center}
\begin{tabular}{llll}
\hline
&ALCOR model & Preliminary data & Ref. \\ \hline\hline
 $h^-$
& 260        & $264 \pm 18$           & STAR [4]   \\
 ${\overline p}^-/{p}^+$
& 0.63       & $0.61 \pm 0.06$              & STAR [4]   \\
 ${\overline \Lambda}/{\Lambda}$
& 0.72       & $0.73 \pm 0.03$              & STAR [4]   \\
 ${\overline \Xi}^+/{\Xi}^-$
& 0.83       & $0.82  \pm 0.08$             & STAR [4]   \\
 $K^+/K^-$
& 1.13       & $1.12 \pm 0.06$              & STAR [4]   \\
 $K^+/\pi^+$
& 0.142      & $0.15 \pm 0.01$              & STAR [13]   \\
 $K^{*+}/h^-$
& 0.077      & 0.065                       & STAR [4]   \\
 ${\overline K}^{*-}/h^-$
& 0.067     & 0.060                        & STAR [4]   \\ \hline
\end{tabular}
\end{center}
\caption{
Hadron production in Au+Au collision at $\sqrt{s}=130$ AGeV
from the ALCOR model and the preliminary experimental data
[4,13]}
\end{table} 

\begin{figure}
\centerline{
\psfig{figure=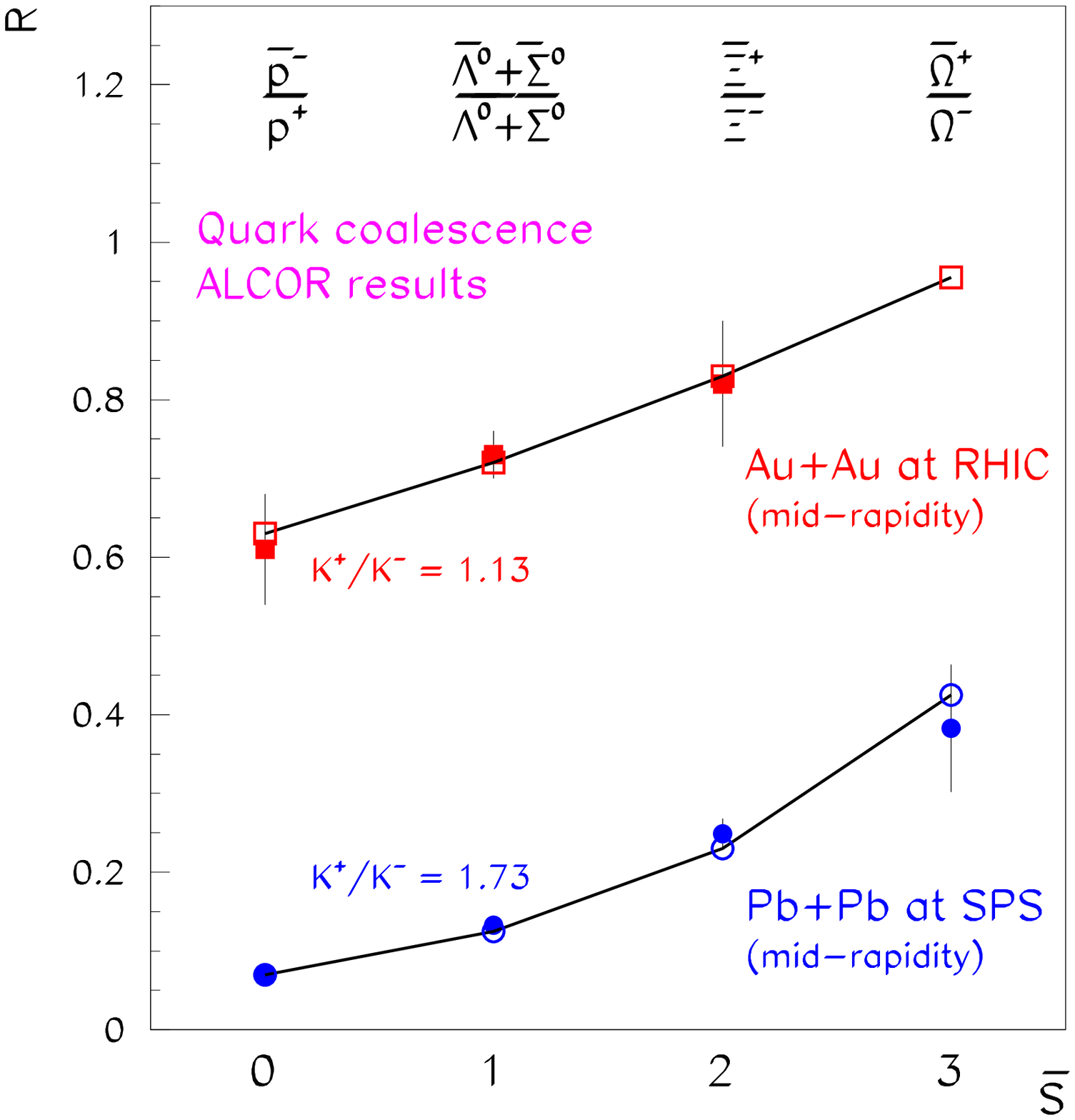,height=4.5in,width=5.0in} }
\vskip -0.05in
\caption[]{
 \label{figure1}
Particle ratios at mid-rapidity 
in Pb+Pb collisions at SPS (open circle)
and in Au+Au collision at RHIC 
(open square). Preliminary experimental data (filled circle and square)
are from Refs.\cite{STAR,STAR2,EXP1,EXP2} 
}
\end{figure}

Investigating particle production in the central rapidity
we repeat our calculation for Pb+Pb collision also at 
CERN SPS energy.
The central rapidity density of the negatively charged hadrons was  
measured by NA49 Collaboration:\cite{NA49h}
\begin{equation}
{\rm d}N_{{\rm h}^-}/{\rm d}y = 195 \pm 15 \ \ .
\label{PbPbh}
\end{equation}
In the ALCOR calculation we use
121 newly produced $u{\overline u}$ and $d{\overline d}$
pairs (this means 53 $s{\overline s}$ pairs)
and assume  17 \% of the total nucleon numbers to be
stopped into the central unit of rapidity (which means 28 proton and
42 neutron). For this case ALCOR model yields 
${\rm d}N_{{\rm h}^-}/{\rm d}y = 195$ in the central rapidity 
and the following baryon/antibaryon ratios:
${\overline p}/p = 0.07$,
${\overline \Lambda}/\Lambda = 0.125$,
${\overline \Xi}/{\Xi} = 0.230$ and 
${\overline \Omega}/{\Omega} = 0.425$.
The obtained ratio $K^+/K^- = 1.73$ fullfils also the expectation,
see eqs.(\ref{SPSrat}-\ref{SPSratk}).

The calculated and the measured antibaryon/baryon ratios at 
SPS and RHIC energies are displayed in Figure 1.

\section{Summary: Quark Matter at CERN SPS and at RHIC}

It seems that the gluonic degrees of freedom are 
liberated in the initial state in Au+Au collisions at RHIC,
as evident from data on jet quenching,\cite{phenix-ismd00,levai-ismd00}
and this initial state seems to be different from that of Pb+Pb
collisions at CERN SPS.\cite{csorgo-qm} As the fireball expands and
cools the gluon
dominated initial state in the $\sqrt{s} = 130$ AGeV Au+Au 
reactions at RHIC  decays to a quark matter kind of state, where 
constituent quark and antiquark degrees of freedom
are dominant.  This quark-antiquark matter
undergoes a sudden hadronization similarly to that at CERN SPS.

The success of the coalescence model to describe the experimental
data at RHIC and SPS strongly supports the validity of the basic 
assumption behind the ALCOR model: before the hadronization,
the basic degrees of freedom of the expanding fireball
are massive (constituent) quarks and antiquarks. 
These elements of the deconfined phase coalesce into final state hadrons
in a sudden process. 

\section*{Acknowledgments}
This research has been supported by the
two Bolyai Fellowships of the Hungarian Academy of Sciences 
(T. Cs. and L. P.), by the Hungarian OTKA grants T024094, T025579, 
T026435, T034269, by the US-Hungarian Joint 
Fund MAKA 652/1998 and by the NWO-OTKA grant N025186.

\end{document}